\documentclass{raa}           
\usepackage{graphicx,times}
\usepackage{natbib}
\usepackage{amssymb,amsmath}
\bibpunct{(}{)}{;}{a}{}{,}

\usepackage[a4paper=true,pagebackref=true]{hyperref}
\hypersetup{pdftitle = The title of my PDF, pdfauthor = My name, pdfsubject= The subject, pdfkeywords = keyword1 keyword2 keyword3} 
\hypersetup{colorlinks = true, linkcolor = green, anchorcolor = red, citecolor = blue, filecolor = red, pagecolor = red, urlcolor = red}

\newcommand{\cmc}{cm$^{-3}$}
\newcommand{\cms}{cm$^{-2}$}

\newcommand{\kms}{km\,s$^{-1}$}

\begin{document}

   \title{Disks and outflows in the S255IR area of high mass star formation from ALMA observations
}

 \volnopage{ {\bf 2018} Vol.\ {\bf X} No. {\bf XX}, 000--000}
   \setcounter{page}{1}

\author{I. Zinchenko\inst{1,2} \and
S.-Y. Liu\inst{3} \and
Y.-N. Su\inst{3} \and
Y. Wang\inst{4}
}
\institute{Institute of Applied Physics of the Russian Academy of Sciences,
           46 Ulyanov~str., 603950 Nizhny Novgorod, Russia\\
              \email{zin@appl.sci-nnov.ru}
              \and
              Lobachevsky State University of Nizhni Novgorod, 23 Gagarin av., Nizhny Novgorod 603950, Russia
         \and
             Institute of Astronomy and Astrophysics, Academia Sinica.
P.O. Box 23-141, Taipei 10617, Taiwan, R.O.C.
\and
Max-Planck-Institut f\"ur Astronomie, K\"onigstuhl 17, D-69117 Heidelberg, Germany
\\
\vs \no
   {\small Received ...; accepted ...}
}

\abstract{We describe the general structure of the well known S255IR high mass star forming region, as revealed by our recent ALMA observations. The data indicate a physical relation of the major clumps SMA1 and SMA2. {The driving source of the extended high velocity well collimated bipolar outflow is not the most pronounced disk-like SMA1 clump harboring a 20~M$_\odot$ young star (S255 NIRS3), as it was assumed earlier. Apparently it is the less evolved SMA2 clump, which drives the outflow and contains a compact rotating structure (probably a disk).} At the same time the SMA1 clump drives another outflow, with a larger opening angle. The molecular line data do not show an outflow from the SMA3 clump (NIRS1), which was suggested by IR studies of this region. 
}

   \authorrunning{I. Zinchenko, et al. }            
   \titlerunning{Disks and outflows in S255IR}  
   \maketitle

%
\section{Introduction}           
\label{sect:intro}
Formation of massive stars attracts an enhanced attention nowadays. Investigations of high mass star forming objects are hampered by the facts that they are rather rare and are located at large distances. New instruments like ALMA make it possible to investigate these objects with a much better angular resolution and sensitivity than earlier. 

The process of high mass star formation is still under debate \citep[e.g.][]{Tan14}. A crucial question is a search for and characterization of disks and outflows in these regions, which may indicate the formation of such stars by disk accretion. The number of known disks around massive (proto)stars is still very limited \citep{Beltran16}.

We have investigated earlier \citep{Zin12,Zin15} the dense core S255IR in the well known high mass star forming complex S255 located at a distance of
$ 1.78^{+0.12}_{-0.11} $~kpc \citep{Burns16}. It consists of several clumps. The most prominent one, S255IR-SMA1, which contains a 20~M$_\odot$ young stellar object S255 NIRS3, shows clearly signs of rotation around an axis approximately {coinciding with} the jet observed here earlier \citep{Wang11,Zin15}. Recently we observed this object with ALMA at a much higher angular resolution and a much better sensitivity than earlier \citep{Zin17a}. One of the findings was a detection of a new methanol maser line \citep{Zin17}. Here we describe the overall structure of the S255IR region as revealed by ALMA.


\section{Observations}
\label{sect:Obs}
The observations reported here were carried out with ALMA in Band 7 at two epochs in 2016 during Cycle 4 under the project 2015.1.00500.s. Details of the observations are given in \cite{Zin17}. Briefly, four spectral windows were observed, centered at around 335.4 GHz, 337.3 GHz, 349.0 GHz, and 346.6 GHz, with bandwidths of 1875.0 MHz, 234.4 MHz, 937.5 MHz, and 1875.0 MHz. 
Data were first calibrated through the ALMA manual or pipeline calibration process. We further employed self-calibration to improve the imaging quality and dynamical range. A continuum image was made using visibilities in the wide low-resolution (continuum) spectral window at 335.4 GHz with obvious spectral contamination removed.
The resulting images achieve an angular resolution of $ 0{\farcs}10\times 0{\farcs}15 $ with Briggs weighting with a robust parameter of 0.5.

\section{Results}
\label{sect:results}
\begin{figure*}
\begin{minipage}[b]{0.68\textwidth}
    \includegraphics[width=\textwidth]{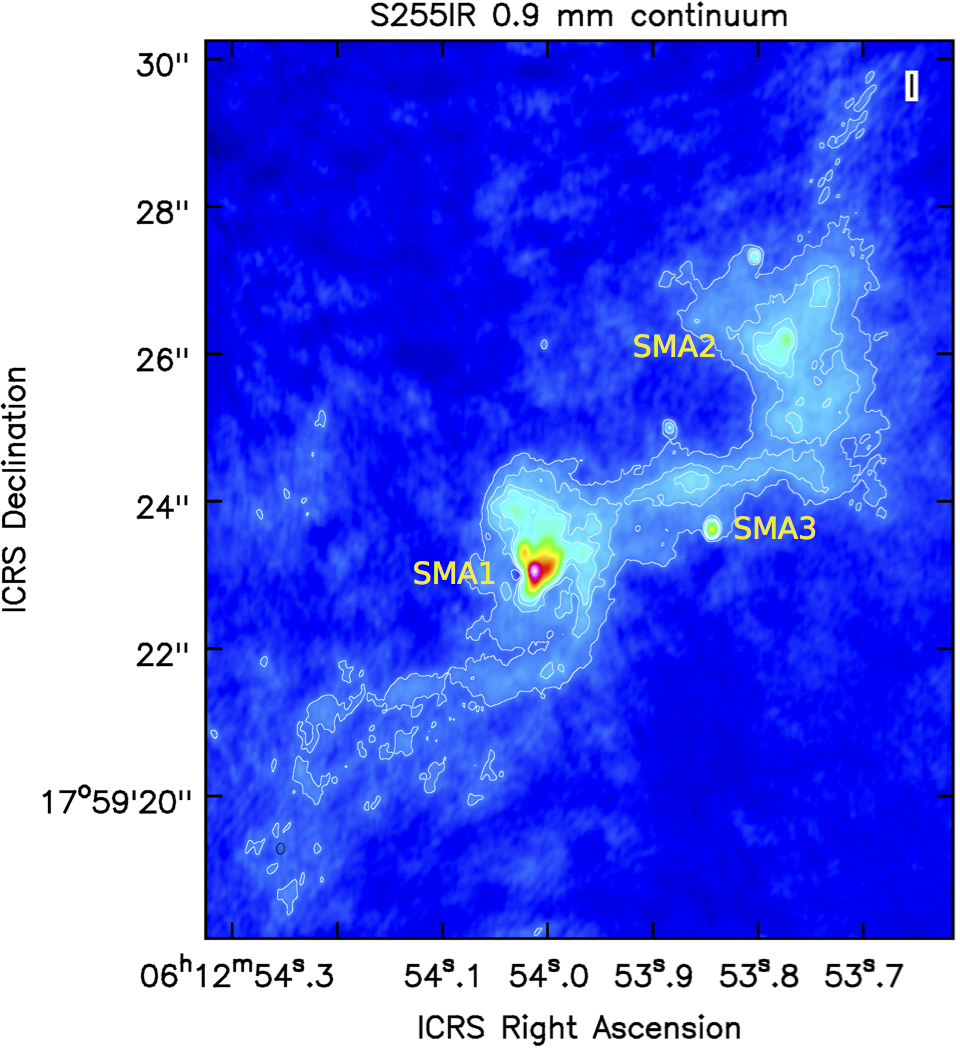}
\end{minipage}
\hfill
\begin{minipage}[b]{0.31\textwidth}
    \caption{The 335~GHz (0.9~mm) continuum image of the S255IR region obtained with ALMA. The SMA1, SMA2 and SMA3 clumps are marked.}
    \label{fig:cont}
\end{minipage}
\end{figure*}

Here we mainly describe the general structure of the region. The details on individual objects will be given elsewhere. The continuum image of the S255IR region at 335~GHz (0.9~mm) is presented in Fig.~\ref{fig:cont}. Previous SMA observations showed here two rather isolated clumps (SMA1 and SMA2) with a hint on a possible presence of the SMA3 clump. 

In the ALMA image all these clumps are clearly seen, as well as some other probably point sources. One can also see that the SMA1 and the SMA2 clumps are located within a structure, which looks like a narrow filament. It is not clear whether SMA3 also belongs to this structure. The apparent length of the filament is about $10^{\prime\prime}$, which corresponds to $ \sim 20000 $~AU. Its width in some places is $\sim 1000$~AU. The continuum brightness temperature in the middle of the filament is about 3~K. 

The SMA1 and SMA2 clumps are rather extended and have complicated morphology while the SMA3 clump is almost point-like. The fluxes of these clumps can be estimated either by integrating their emission over some area or by fitting 2D Gaussians. Both approaches have shortcomings because in the first case it is not clear how to select the integration area while the second one can be misleading due to a complicated source structure. With these caveats in mind we obtain the continuum flux densities $ \sim 1 $~Jy for SMA1, $ \sim 0.5 $~Jy for SMA2, and $ \sim 50 $~mJy for SMA3. 

\begin{figure*}
\centering
    \includegraphics[width=0.9\textwidth]{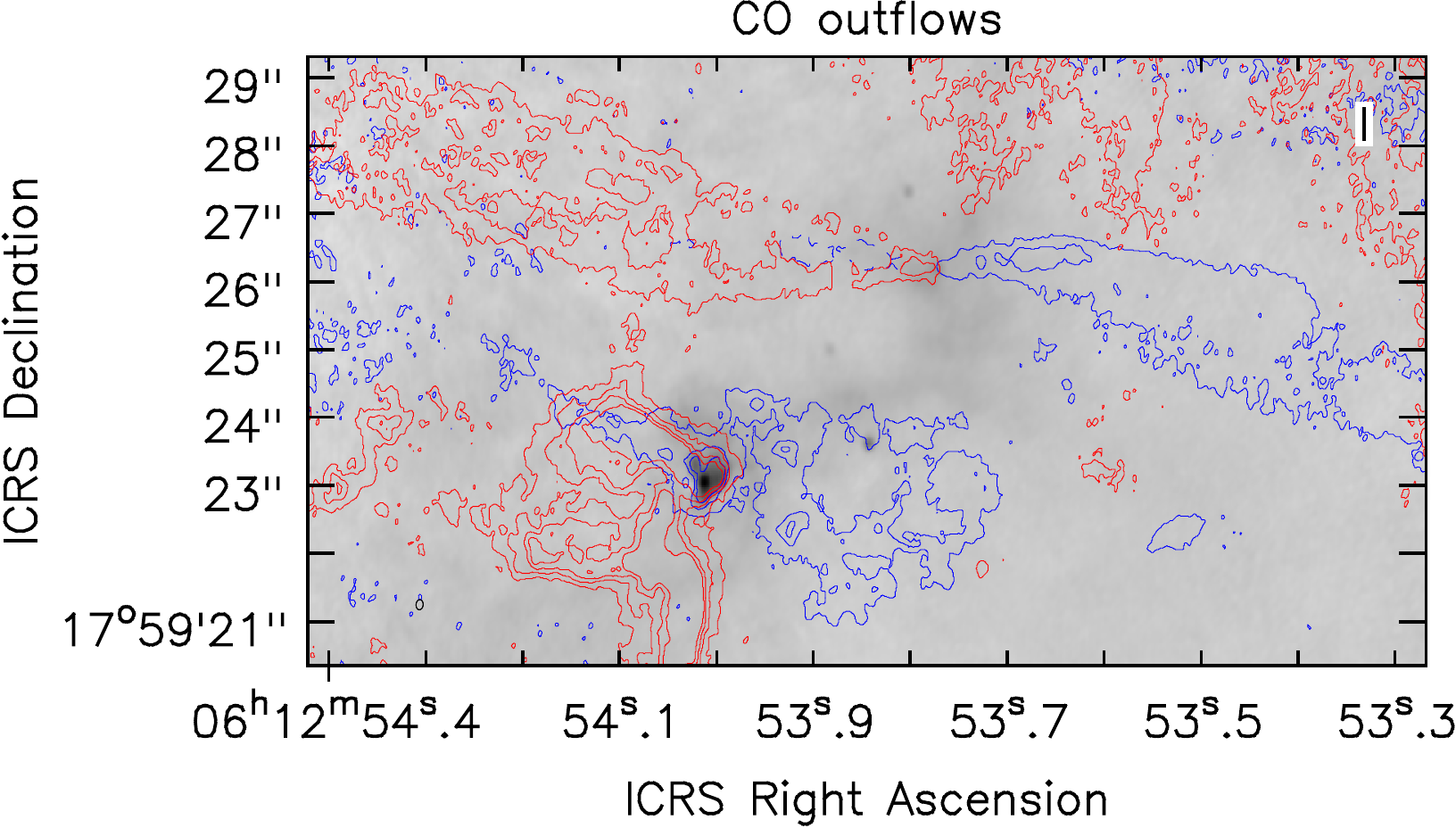}
    \caption{Maps of the red- and blue-shifted high velocity CO(3--2) emission in the S255IR area, overlaid on the continuum image. The blue-shifted emission is integrated in the velocity range from --67~\kms\ to --16~\kms, the red-shifted emission is from 20~\kms\ to 90~\kms.}
    \label{fig:outflows}
\end{figure*}

Previous studies have shown a presence of high velocity outflow(s) here.
Now we use the CO(3--2) data to investigate them. The maps of the red- and blue-shifted high velocity CO(3--2) emission are presented in Fig.~\ref{fig:outflows}, overlaid on the continuum image. 
This figure shows clearly two outflows originating at the SMA1 and SMA2 clumps. The outflow from SMA2 is well collimated and appear much more extended than the outflow from SMA1. It is also characterized by larger velocities. It is worth mentioning that this outflow is apparently bended. The outflow from SMA1 has a wider opening angle and is more compact. There is no sign of any outflow from SMA3.

\begin{figure*}
\begin{minipage}[b]{0.55\textwidth}
    \includegraphics[width=\textwidth]{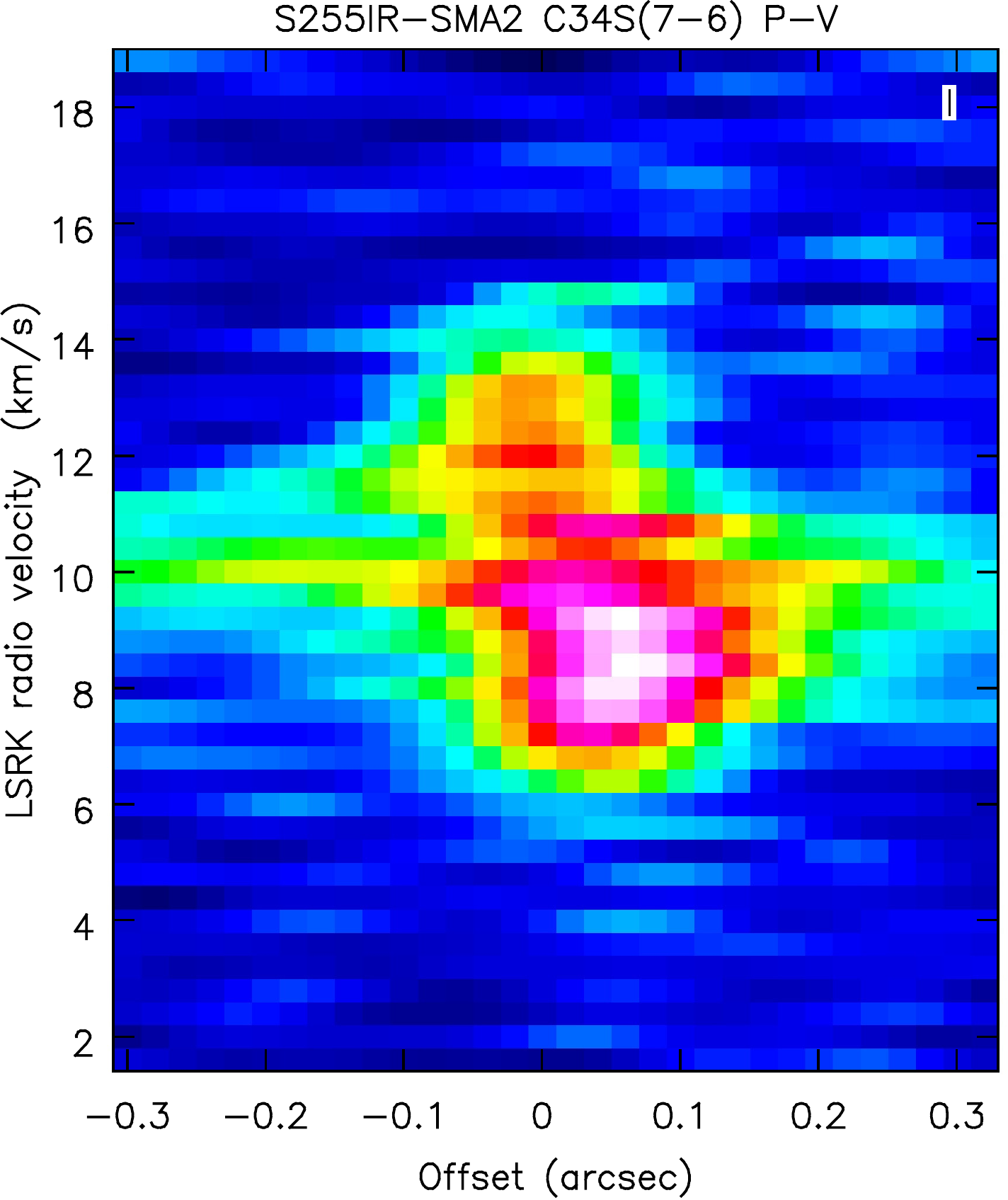}
\end{minipage}
\hfill
\begin{minipage}[b]{0.44\textwidth}
    \caption{The position-velocity diagram for the S255IR-SMA2 clump in the C$^{34}$S(7--6) line at the position angle of 150$^\circ$.}
    \label{fig:sma2-pv}
\end{minipage}
\end{figure*}

It is reasonable to expect that the impressive outflow from SMA2 is associated with a disk. Earlier no sign of rotation has been reported for SMA2. In our ALMA data several molecular lines are seen toward this clump, in particular a rather strong emission in the C$^{34}$S(7--6) line. The position-velocity diagram in this line is presented in Fig.~\ref{fig:sma2-pv}. It clearly shows a Keplerian-like rotation of the clump. A crude estimate of the central mass gives a value $ M\sin^2i \sim 1 $~M$_\odot$, where $i$ is the unknown inclination angle.

\section{Discussion}
\label{sect:discussion}
The picture of the S255IR region revealed by the ALMA data is significantly different from that one assumed earlier. The previous SMA and IRAM-30m data were interpreted as an evidence of the outflow from the SMA1 clump as a driving source \citep{Wang11,Zin15} although the the SMA data alone hint at the SMA2 as a possible source of the outflow. The new data do not deny an outflow from SMA1 but they show that this outflow as seen by ALMA is much less extended and not so collimated, as assumed by the previous studies. An extended and highly collimated outflow originates from the SMA2 clump, which also represents a rotating core. This core is not associated with any known IR or maser source. Probably this is a disk around a protostar of a lower mass than that in SMA1. The fact that the outflow is bended may indicate a precession of this disk.

There are evidences for episodic ejections from SMA1 \citep{Zin15,Burns16}. It can be that now we see only a part of the more extended outflow from this massive clump.
The outflows from the SMA1 and SMA2 are almost parallel and it is not excluded that they can merge at larger scales.

The SMA1 and SMA2 clumps are located within a very thin and apparently dense filamentary structure. Its width is more by an order of magnitude smaller than typical values for interstellar filaments \citep{Andre14}. The gas column density, roughly estimated from the continuum brightness, is $ \sim 10^{24} $~\cms, which implies a volume density of $ \sim 10^{8} $~\cmc. A high density is confirmed by observations of molecular transitions with a high critical density, like C$^{34}$S(7--6).

The SMA3 clump looks isolated and very compact. It is associated with the S255 NIRS1 source whose luminosity corresponds to a B3 star \citep{Simpson09}. The measured flux density when compared with the measurements at 225~GHz \citep{Wang11} implies a spectral slope close to that of black body radiation. It is not excluded that the previous SMA observations performed with a much larger beam are contaminated with a surrounding emission. The size of this source is not significantly larger than our beam size, which corresponds to about 200~AU. The IR data indicate a probable outflow from this object \citep{Simpson09}. However our CO data do not show any sign of molecular outflow.

\section{Conclusions}
From our ALMA data a new picture of the well known S255IR star forming region emerges. It is more complicated than assumed earlier, contains unusual objects and several centers of star formation activity, which deserve further detailed investigations.

\normalem
\begin{acknowledgements}
We are grateful to the anonymous referee for the helpful comments.
This study was supported by the Russian Foundation for Basic Research (grant No. 15-02-06098) in the part of the preparation of the observations and initial data reduction and by the Russian Science Foundation (grant No. 17-12-01256) in the part of the spectral data analysis. Y.W. acknowledges support from the European Research Council under the Horizon 2020 Framework Program via the ERC Consolidator Grant CSF-648505. 
This paper makes use of the following ALMA data: ADS/JAO.ALMA\#2015.1.00500.S. ALMA is a partnership of ESO (representing its member states), NSF (USA) and NINS (Japan), together with NRC (Canada), MOST and ASIAA (Taiwan), and KASI (Republic of Korea), in cooperation with the Republic of Chile. The Joint ALMA Observatory is operated by ESO, AUI/NRAO and NAOJ.

\end{acknowledgements}
  
\bibliographystyle{raa}
\bibliography{0257}

\end{document}